\documentclass[letterpaper,12pt,leqno]{article}
\usepackage{paper}
\usepackage{adjustbox}
\hypersetup{pdftitle={Paper Title}}
\newcommand{\bib}{bibliography.bib}

\begin{document}
\title{\Large Two Level Nested and Sequential Logit}
\author{Davide Luparello \thanks{\href{mailto: dluparello@psu.edu}{dluparello@psu.edu}. I am grateful to Conor Ryan for his guidance through the theoretical complexities of the two-level nested logit framework.}}
\affil{{\normalsize \it The Pennsylvania State University} }
\date{\today\\
 \textcolor{blue}{{\href{https://drive.google.com/file/d/1fI7ed_aKt3Cj0aWMOZnQPMJjqPz4LDG0/view?usp=sharing}{\small[click here for the most updated version]}}}}                   
\begin{titlepage}\maketitle
\begin{abstract}
This technical note provides comprehensive derivations of fundamental equations in two-level nested and sequential logit models for analyzing hierarchical choice structures. We present derivations of the \cite{berry1994estimating} inversion formula, nested inclusive values computation, and multi-level market share equations, complementing existing literature. While conceptually distinct, nested and sequential logit models share mathematical similarities and, under specific distributional assumptions, yield identical inversion formulas—offering valuable analytical insights. These notes serve as a practical reference for researchers implementing multi-level discrete choice models in empirical applications, particularly in industrial organization and demand estimation contexts, and complement \cite{mansley2019notes}.
\end{abstract}

\end{titlepage}

\doublespacing
\section{Two Level Nested Logit}

\subsection{Utility Specification}
In a two-level nested logit framework, the indirect utility of consumer $i$ selecting product $j$, which belongs to subgroup $h$ within group $g$, is modeled as:
\begin{equation}
    U_{ijhg} = \delta_j + \zeta_i^{g} + (1 - \sigma_2) \zeta_i^{h} + (1 - \sigma_1) \epsilon_{ij}.
\end{equation}
Here, $\delta_j$ denotes the mean utility derived from product $j$, which is common across consumers. The parameters $\sigma_1$ and $\sigma_2$ are nesting parameters capturing the substitutability of products within subgroups and across subgroups, respectively. Higher values of these parameters indicate stronger substitutability within or between nests. Additionally, the model incorporates several stochastic terms: $\zeta_i^{g}$ is a group-level taste shock specific to consumer $i$, $\zeta_i^{h}$ captures the subgroup-level preferences of consumer $i$, and $\epsilon_{ij}$ represents the idiosyncratic, product-specific preference shock for consumer $i$. These stochastic components account for unobserved variations in consumer preferences and decision-making.\\
\newline
All random components—$\zeta_i^{g}$, $\zeta_i^{h}$, and $\epsilon_{ij}$—are assumed to be mutually independent. Specifically, $\epsilon_{ij}$ follows a standard Type-I extreme value distribution, while $\zeta_i^{g}$ and $\zeta_i^{h}$ have unique distributions designed such that the following composite error terms:
\begin{align}
&\zeta_i^{g} + (1 - \sigma_2)\zeta_i^{h} + (1 - \sigma_1)\epsilon_{ij}, \label{eq:assg} \\
&\zeta_i^{h} + \frac{(1 - \sigma_1)}{(1 - \sigma_2)}\epsilon_{ij}, \label{eq:assh}
\end{align}
are both distributed as Type-I extreme value. The first composite term has a scale parameter $(1-\sigma_1)$ and a location parameter $m_1$, and the second composite term has a scale parameter $\frac{(1-\sigma_1)}{(1-\sigma_2)}$ and a location parameter $m_2$\footnote{This setup aligns with Theorem 2.1 in \cite{cardell1997variance}, which also imposes the restriction $0<\sigma_2<\sigma_1<1$. Additional intuition can be found in the discussion following Lemma 2.2 of the same paper.}.\\

\subsection{Hierarchy of Conditional Choice Probabilities}
Due to the hierarchical nature of the model, choice probabilities are derived through sequential conditioning, starting from the lowest level (individual product choices) and proceeding upward to higher levels (subgroup and group-level decisions).

\subsubsection{Product Choice Probability}
Conditional on selecting group $g$ and subgroup $h$, agent $i$ chooses product $j$ with probability:
\begin{equation}
P(j|h,g) = \Pr\left\{ U_{ijhg} \geq U_{ij'hg} \; \forall \; j' \in J_{hg},\;j'\neq j\;|h,g \;\text{chosen} \right\},
\end{equation}
where $J_{hg}$ denotes the set of products offered in subgroup $h$ of group $g$. By substituting the decomposed utilities and simplifying, this conditional probability becomes:
\begin{equation}
P(j|h,g)=\Pr\left\{\epsilon_{ij} \geq \frac{\delta_{j'} - \delta_j}{(1-\sigma_1)} + \epsilon_{ij'} \;\forall \; j' \in J_{hg},\;j'\neq j\right\}.
\end{equation}
Leveraging the distributional properties of $\epsilon_{ij}$, which is standard Type-I extreme value, the conditional choice probability simplifies further to:
\begin{equation}
P(j|h,g) = \frac{\exp\left(\frac{\delta_{j}}{1-\sigma_1}\right)}{\sum_{j' \in J_{hg}} \exp\left(\frac{\delta_{j'}}{1-\sigma_1}\right)}.
\end{equation}
Defining the inclusive value as:
\begin{equation}
I_{hg} = (1-\sigma_1)\log\sum_{j' \in J_{hg}} \exp\left(\frac{\delta_{j'}}{1-\sigma_1}\right),
\end{equation}
the conditional choice probability can then be expressed succinctly as:
\begin{equation}
P(j|h,g) = \frac{\exp\left(\frac{\delta_{j}}{1-\sigma_1}\right)}{\exp\left(\frac{I_{hg}}{1-\sigma_1}\right)}.
\end{equation}

\subsubsection{Product Subgroup Choice Probability}
Given that agent $i$ selects group $g$, the probability of choosing a product in subgroup $h$ within group $g$ is expressed as:
\begin{equation}
P(h|g)=\Pr\left\{ \max_{j\in J_{hg}} U_{ijhg} \geq \max_{j'\in J_{h'g}} U_{ij'h'g} \quad \forall\,h'\in H_g,\,h'\neq h\,|\,g \;\text{chosen}\right\},
\end{equation}
where $H_{g}$ denotes the set of subgroups within group $g$.\\
\newline
Substituting the decomposed form of the indirect utilities and rearranging yields:
\begin{equation}
P(h|g)=\Pr\left\{ \max_{j\in J_{hg}}\left[\frac{\delta_j}{1-\sigma_2} + \zeta_i^h + \frac{1-\sigma_1}{1-\sigma_2}\epsilon_{ij}\right] \geq \max_{j'\in J_{h'g}}\left[\frac{\delta_{j'}}{1-\sigma_2} + \zeta_i^{h'} + \frac{1-\sigma_1}{1-\sigma_2}\epsilon_{ij'}\right] \forall\,h'\in H_g,\,h'\neq h\right\}.
\end{equation}
Since the composite term $\zeta_i^h + \frac{1-\sigma_1}{1-\sigma_2}\epsilon_{ij} - m_1$ is Type-I extreme value distributed with scale parameter $\frac{1-\sigma_1}{1-\sigma_2}$ and location parameter $0$, by the maximum stability property of the extreme value distribution, we obtain:
\begin{equation}
\max_{j\in J_{hg}}\left[\frac{\delta_j}{1-\sigma_2} + \zeta_i^h + \frac{1-\sigma_1}{1-\sigma_2}\epsilon_{ij} - m_1\right]
\end{equation}
being itself Type-I extreme value distributed with unit scale and location parameter:
\begin{equation}
\eta_h = \frac{1-\sigma_1}{1-\sigma_2}\log\sum_{j\in J_{hg}}\exp\left(\frac{\delta_j}{1-\sigma_1}\right) = \frac{I_{hg}}{1-\sigma_2}.
\end{equation}
Thus, the conditional probability simplifies to:
\begin{equation}
P(h|g)=\Pr\left\{\tilde{\epsilon}_{ih} \geq (\eta_{h'} - \eta_h) + \tilde{\epsilon}_{ih'}\quad \forall\,h'\in H_g,\,h'\neq h\right\},
\end{equation}
where the normalized composite error term is defined as:
\begin{equation}
\tilde{\epsilon}_{ih}=\max_{j\in J_{hg}}\left[\frac{\delta_j}{1-\sigma_2} + \zeta_i^h + \frac{1-\sigma_1}{1-\sigma_2}\epsilon_{ij}-m_1\right]-\eta_h.
\end{equation}
Applying properties of Type-I extreme value distributions, we have:
\begin{equation}
P(h|g)=\frac{\exp\left(\frac{I_{hg}}{1-\sigma_2}\right)}{\sum_{h'\in H_g}\exp\left(\frac{I_{h'g}}{1-\sigma_2}\right)}.
\end{equation}
Expressing explicitly in terms of product-level utilities, we obtain:
\begin{equation}
P(h|g)=\frac{\left(\sum_{j\in J_{hg}}\exp\left(\frac{\delta_j}{1-\sigma_1}\right)\right)^{\frac{1-\sigma_1}{1-\sigma_2}}}{\sum_{h'\in H_g}\left(\sum_{j\in J_{h'g}}\exp\left(\frac{\delta_j}{1-\sigma_1}\right)\right)^{\frac{1-\sigma_1}{1-\sigma_2}}}.
\end{equation}
Defining the inclusive value at the group level as:
\begin{equation}
I_{g}=(1-\sigma_2)\log\sum_{h\in H_g}\exp\left(\frac{I_{hg}}{1-\sigma_2}\right),
\end{equation}
the subgroup choice probability can be succinctly written as:
\begin{equation}
P(h|g)=\frac{\exp\left(\frac{I_{hg}}{1-\sigma_2}\right)}{\exp\left(\frac{I_{g}}{1-\sigma_2}\right)}.
\end{equation}

\subsubsection{Product Group Choice Probability}

The probability of consumer $i$ selecting a product within group $g$ is expressed as:
\begin{equation}
P(g)=\Pr\left\{ \max_{h \in H_g}\max_{j \in J_{hg}} U_{ijhg} \geq \max_{h' \in H_{g'}}\max_{j' \in J_{h'g'}} U_{ij'h'g'} \quad \forall\, g' \neq g \right\},
\end{equation}
Substituting and rearranging the decomposed utility expressions, this probability becomes:
\begin{equation}
\begin{aligned}
P(g)=\Pr\Big\{&\max_{h \in H_g}\max_{j \in J_{hg}}\left[ \delta_j + \zeta_i^{g} + (1-\sigma_2)\zeta_i^{h} + (1-\sigma_1)\epsilon_{ij}\right] \geq \\
&\max_{h' \in H_{g'}}\max_{j' \in J_{h'g'}}\left[ \delta_{j'} + \zeta_i^{g'} + (1-\sigma_2)\zeta_i^{h'} + (1-\sigma_1)\epsilon_{ij'}\right]\quad \forall\, g' \neq g \Big\}.
\end{aligned}
\end{equation}
Given that the composite error term $\zeta_i^{g} + (1-\sigma_2)\zeta_i^{h} + (1-\sigma_1)\epsilon_{ij}-m_2$ is Type-I extreme value distributed with scale parameter $(1-\sigma_1)$ and location parameter $0$, the term:
\begin{equation}
\max_{j \in J_{hg}}\left[ \delta_j + \zeta_i^{g} + (1-\sigma_2)\zeta_i^{h} + (1-\sigma_1)\epsilon_{ij}-m_2\right]
\end{equation}
is also Type-I extreme value distributed with scale parameter $1$ and location parameter:
\begin{equation}
\eta_{h2} =(1-\sigma_1)\log\sum_{j\in J_{hg}}\exp\left(\frac{\delta_j}{1-\sigma_1}\right)=I_{hg}.
\end{equation}
Consequently, the group-level choice probability simplifies to:
\begin{equation}
P(g)=\Pr\left\{ \max_{h \in H_g}[\tilde{\epsilon}_{ih2} +\eta_{h2}]\geq \max_{h' \in H_{g'}}[\tilde{\epsilon}_{ih'2} +\eta_{h'2}] \quad \forall\, g' \neq g \right\},
\end{equation}
where the normalized composite error term is defined as:
\begin{equation}
\tilde{\epsilon}_{ih2}= \max_{j \in J_{hg}}\left[ \delta_j + \zeta_i^{g} + (1-\sigma_2)\zeta_i^{h} + (1-\sigma_1)\epsilon_{ij}-m_2\right]-\eta_{h2}.
\end{equation}
Since $\tilde{\epsilon}_{ih2}$ is Type-I extreme value distributed with scale parameter $(1-\sigma_2)$ and zero location, it follows that the term:
\begin{equation}
\max_{h \in H_g}[\tilde{\epsilon}_{ih2} +\eta_{h2}]
\end{equation}
is Type-I extreme value distributed with unit scale and location parameter:
\begin{equation}
\eta_{g} =(1-\sigma_2)\log\sum_{h\in H_{g}}\exp\left(\frac{I_{hg}}{1-\sigma_2}\right)=I_g.
\end{equation}
Therefore, the group-level choice probability can be expressed succinctly as:
\begin{equation}
P(g)=\frac{\exp(I_g)}{\sum_{g'}\exp(I_{g'})}.
\end{equation}
Explicitly substituting for the inclusive values, the probability becomes:
\begin{equation}
P(g) = \frac{\left(\sum_{h \in H_g} \left(\sum_{j \in J_{hg}} \exp\left(\frac{\delta_{j}}{1-\sigma_1}\right)\right)^{\frac{1-\sigma_1}{1-\sigma_2}}\right)^{(1-\sigma_2)}}{\sum_{g'}\left(\sum_{h \in H_{g'}} \left(\sum_{j \in J_{h'g'}} \exp\left(\frac{\delta_{j}}{1-\sigma_1}\right)\right)^{\frac{1-\sigma_1}{1-\sigma_2}}\right)^{(1-\sigma_2)}}.
\end{equation}
Finally, defining the top-level inclusive value as:
\begin{equation}
I = \log\sum_{g'} \exp\left(I_{g'}\right),
\end{equation}
the group choice probability can be succinctly rewritten as:
\begin{equation}
P(g) = \frac{\exp\left(I_g\right)}{\exp\left(I\right)}.
\end{equation}

\subsection{Joint Choice Probability and Berry (1994) Inversion}

The hierarchical structure of the nested logit model allows the joint probability of choosing a product $j$ in subgroup $h$ of group $g$ to be factorized into a product of conditional probabilities. Specifically, the joint probability can be expressed as:
\begin{equation}
s_{jhg} = P(j,h,g) = P(j|h,g) \cdot P(h|g) \cdot P(g),
\end{equation}
where $s_{jhg}$ denotes the market share of product $j$ within subgroup $h$ of group $g$.\\
\newline
To derive explicit product demand equations, we apply the nested logit inversion methodology introduced by \cite{berry1994estimating}. Normalizing the inclusive value of the outside option (representing non-participation) to zero, i.e., $I_{0} = 0$, the market share of the outside option is obtained as:
\begin{equation}
s_{0} = P(g=0) = \frac{1}{\sum_{g'} \exp(I_{g'})}.
\end{equation}
Consequently, the ratio of the product's market share to the outside option share can be formulated as:
\begin{equation}
\frac{s_{jhg}}{s_{0}} = \frac{\exp\left(\frac{\delta_{j}}{1-\sigma_1}\right)}{\sum_{j' \in J_{hg}} \exp\left(\frac{\delta_{j'}}{1-\sigma_1}\right)} \cdot \frac{\left(\sum_{j \in J_{hg}} \exp\left(\frac{\delta_{j}}{1-\sigma_1}\right)\right)^{\frac{1-\sigma_1}{1-\sigma_2}}}{\sum_{h'} \left(\sum_{j \in J_{h'g}} \exp\left(\frac{\delta_{j}}{1-\sigma_1}\right)\right)^{\frac{1-\sigma_1}{1-\sigma_2}}} \cdot \left(\sum_{h}  \left(\sum_{j \in J_{hg}} \exp\left(\frac{\delta_{j}}{1-\sigma_1}\right)\right)^{\frac{1-\sigma_1}{1-\sigma_2}}\right)^{(1-\sigma_2)}.
\end{equation}
By algebraic simplification and taking logarithms, this ratio simplifies elegantly to the familiar \cite{berry1994estimating} inversion formula:
\begin{equation}
\log\left(\frac{s_{jhg}}{s_{0}}\right) = \delta_{j} + \sigma_1\log s_{j|hg} + \sigma_2\log s_{h|g},
\end{equation}
where $s_{j|hg} = P(j|h,g)$ denotes the conditional market share of product $j$ within subgroup $h$ of group $g$, and $s_{h|g} = P(h|g)$ denotes the conditional market share of subgroup $h$ within group $g$. 

\section{Two Level Sequential Logit}

\subsection{Utility Specification}
The utility function for agent $i$ selecting product $j$ in subgroup $h$ of group $g$ is specified as:
\begin{equation}
U_{ijhg} = \delta_j + \zeta^g_{i} + (1-\sigma_2)\zeta^h_{i} + (1-\sigma_1)\epsilon_{ij}
\end{equation}
In this specification, $\delta_j$ represents the mean utility derived from product $j$, which remains constant across all consumers. The stochastic components establish a sequential choice structure that systematically captures preference heterogeneity. The random terms $\epsilon_{ij}$, $\zeta^h_{i}$, and $\zeta^g_{i}$ are independently distributed as standard Type-I extreme value distributions\footnote{This distributional assumption fundamentally differentiates the sequential logit from the nested logit framework. See \cite{cardell1997variance} for the distributional requirements in nested logit models.}.\\
\newline
The nesting parameters $\sigma_1$ and $\sigma_2$ (where $0 \leq \sigma_1, \sigma_2 \leq 1$) govern the information structure and choice dynamics across hierarchical decision levels. Specifically, $\sigma_1$ modulates the variance of utility shocks at the product level relative to information available when selecting a subgroup, while $\sigma_2$ performs an analogous function at the subgroup level relative to information available when selecting a group. As these parameters approach unity, choices become increasingly deterministic at their respective levels, reflecting greater information precision during the sequential decision process. This interpretation stands in contrast to the nested logit framework, where these parameters instead characterize the correlation of preferences within and across subgroups, leading to fundamentally different behavioral implications.\\
\newline
The sequential nature of consumer decision-making unfolds in a specific temporal order: initially, consumer $i$ selects group $g$ based solely on knowledge of the group-specific shock $\zeta^g_i$ and the deterministic utility components. Subsequently, upon choosing group $g$, the consumer observes the subgroup-specific shock $\zeta^h_i$ and utilizes this additional information to select subgroup $h$ within the chosen group. Finally, after committing to subgroup $h$, the consumer observes the product-specific shock $\epsilon_{ij}$, which guides the ultimate product selection from the available alternatives within that subgroup. This progressive revelation of information characterizes the sequential decision process and distinguishes it from simultaneous choice models.

\subsection{Hierarchy of Conditional Choice Probabilities}
The model's two-level nested structure necessitates deriving choice probabilities through sequential conditioning, proceeding from the lowest level (product choice) to the highest (product group decision).

\subsubsection{Product Choice Probability}
Conditional on selecting group $g$ and subgroup $h$, agent $i$ chooses product $j$ with probability:
\begin{equation}
P(j|h,g) = \Pr\Big\{ U_{ijhg} \geq U_{ij'hg} \quad \forall \; j' \in J_{hg},\;j'\neq j\;|\; h,g \;\text{chosen} \Big\}
\end{equation}
where $J_{hg}$ denotes the set of products offered in subgroup $h$ of group $g$.\\
\newline
By substituting the decomposed utilities on both sides and simplifying, I obtain the conditional choice probability:
\begin{equation}
P(j|h,g)=\Pr\left\{\epsilon_{ij} \geq \frac{\delta_{j'} - \delta_j}{(1-\sigma_1)} \epsilon_{ij'} \quad \forall \; j' \in J_{hg},\;j'\neq j\;|\; h,g \;\text{chosen}\right\}
\end{equation}
Applying distributional properties of extreme value variables yields:
\begin{equation}
\label{eq:product_choice}
P(j|h,g) = \frac{\exp\left(\frac{\delta_{j}}{1-\sigma_1}\right)}{\sum_{j' \in J_{hg}} \exp\left(\frac{\delta_{j'}}{1-\sigma_1}\right)}
\end{equation}

\subsubsection{Product Subgroup Choice Probability}

Consumer $i$ selects subgroup $h$ within group $g$ if and only if it provides the highest expected utility among all available subgroups within that group:
\begin{equation}
V_{ihg} \geq V_{ih'g} \quad \forall h'\in H_g,\; h' \neq h
\end{equation}
where $H_{g}$ denotes the set of subgroups within group $g$.\\
\newline
The expected maximum utility attainable by consumer $i$ when selecting among products in subgroup $h$ of group $g$, after observing group and subgroup shocks but before observing the product-specific shock $\epsilon_{ij}$, is expressed as:
\begin{equation}
\begin{aligned}
    V_{ihg} &= \mathbf{E}_\epsilon\left[\max_{j \in J_{hg}} U_{ijhg}\;|\;\zeta^g_i, \zeta^h_i\right] \\
    &= \mathbf{E}_\epsilon\left[\max_{j \in J_{hg}} \left\{\delta_{j}+ \zeta^g_{i}+(1-\sigma_2)\zeta^h_{i} + (1-\sigma_1)\epsilon_{ij}\right\}\;|\;\zeta^g_i, \zeta^h_i\right]
\end{aligned}
\end{equation}
At this decision stage, the components $\zeta^g_{i}$ and $\zeta^h_{i}$ are already observed by the consumer. Consequently, the composite term $[\zeta^g_{i}+(1-\sigma_2)\zeta^h_{i} + (1-\sigma_1)\epsilon_{ij}]$ follows a Type-I extreme value distribution with scale parameter $(1-\sigma_1)$ and location parameter $[\zeta^g_{i}+(1-\sigma_2)\zeta^h_{i}]$.\\
\newline
Applying \cite{mcfadden1981econometric} seminal result on the expected maximum of extreme value random variables:
\begin{equation}
V_{ihg} = I_{hg} + \zeta^g_{i}+(1-\sigma_2)\zeta^h_{i} + C
\end{equation}
where $C$ is a constant derived from Euler's constant, and $I_{hg}$ represents the inclusive value of subgroup $h$ in group $g$:
\begin{equation}
    I_{hg}=(1-\sigma_1)\ln\sum_{j \in J_{hg}} \exp\left(\frac{\delta_{j}}{1-\sigma_1}\right)
\end{equation}
The probability that consumer $i$ selects subgroup $h$ conditional on having chosen group $g$ is:
\begin{equation}
P(h|g) = \Pr\Big\{ V_{ihg} \geq V_{ih'g} \quad \forall \; h' \in H_{g},\;h'\neq h\;|\; g \;\text{chosen} \Big\}
\end{equation}
Substituting the expression for $V_{ihg}$ and simplifying:
\begin{equation}
P(h|g)=\Pr\left\{\zeta^h_{i} \geq \frac{I_{h'g} - I_{hg}}{(1-\sigma_2)} +\zeta^{h'}_{i}\quad \forall \; h' \in H_{g},\;h'\neq h\;|\; g \;\text{chosen}\right\}
\end{equation}
Since $\zeta^h_{i}$ follows a Type-I extreme value distribution, this conditional choice probability resolves to the familiar logit form:
\begin{equation}
P(h|g) = \frac{\exp\left(\frac{I_{hg}}{1-\sigma_2}\right)}{\sum_{h' \in H_g} \exp\left(\frac{I_{h'g}}{1-\sigma_2}\right)}
\end{equation}
Substituting the explicit formula for the inclusive value $I_{hg}$ yields the comprehensive expression for the subgroup choice probability:
\begin{equation}
\label{eq:subgroup}
P(h|g) = \frac{\left(\sum_{j \in J_{hg}} \exp\left(\frac{\delta_{j}}{1-\sigma_1}\right)\right)^{\frac{1-\sigma_1}{1-\sigma_2}}}{\sum_{h' \in H_g} \left(\sum_{j \in J_{h'g}} \exp\left(\frac{\delta_{j}}{1-\sigma_1}\right)\right)^{\frac{1-\sigma_1}{1-\sigma_2}}}
\end{equation}
This expression characterizes how consumers select among subgroups based on the expected utility derived from the products contained within each subgroup, modulated by the nesting parameters that govern the sequential information structure.

\subsubsection{Product Group Choice Probability}

At the highest level of the sequential structure, consumers select among product groups by evaluating the expected maximum attainable utility across all possible choices. This decision process incorporates information from lower-level choices through inclusive values that systematically aggregate utilities across both products and product subgroups.\\
\newline
Under utility maximization, consumer $i$ selects group $g$ if and only if it yields expected utility at least as high as any alternative group:
\begin{equation}
V_{ig} \geq V_{ig'} \quad \forall g' \neq g
\end{equation}
The expected maximum utility attainable by consumer $i$ from selecting group $g$, before observing the subgroup-specific shock $\zeta^h_i$ but after observing the group-specific shock $\zeta^g_i$, is expressed as:
\begin{equation}
\begin{aligned}
V_{ig} &= \mathbf{E}_{\zeta^h}\left[\max_{h \in H_g} \mathbf{E}_\epsilon\left[\max_{j \in J_{hg}} U_{ijhg}\;|\;\zeta^g_i, \zeta^h_i\right]\;|\;\zeta^g_i\right]\\
&= \mathbf{E}_{\zeta^h}\left[\max_{h \in H_g} \mathbf{E}_\epsilon\left[\max_{j \in J_{hg}} \left\{\delta_{j} + \zeta^g_{i}+(1-\sigma_2)\zeta^h_{i}+ (1-\sigma_1)\epsilon_{ij}\right\}\;|\;\zeta^g_i, \zeta^h_i\right]\;|\;\zeta^g_i\right]
\end{aligned}
\end{equation}
From the subgroup choice derivation, we can reformulate this expression using the previously defined $V_{ihg}$:
\begin{equation}
V_{ig} = \mathbf{E}_{\zeta^h}\left[\max_{h \in H_g} V_{ihg}\;|\;\zeta^g_i\right] = \mathbf{E}_{\zeta^h}\left[\max_{h \in H_g}\left\{ I_{hg} + \zeta^g_{i}+(1-\sigma_2)\zeta^h_{i} + C\right\}\;|\;\zeta^g_i\right]
\end{equation}
At this decision stage, consumer $i$ has observed the component $\zeta^g_i$ but not yet $\zeta^h_i$. Consequently, the composite term $[I_{hg} + \zeta^g_{i}+(1-\sigma_2)\zeta^h_{i}]$ follows a Type-I extreme value distribution with scale parameter $(1-\sigma_2)$ and location parameter $[I_{hg} + \zeta^g_{i}]$.\\
\newline
Applying \cite{mcfadden1981econometric} result for the expected maximum of extreme value random variables:
\begin{equation}
V_{ig} = I_g + \zeta_i^g + C'
\end{equation}
where $C'$ consolidates the constants from the Euler terms, and $I_g$ represents the group-level inclusive value: 
\begin{equation}
    I_g = (1-\sigma_2)\ln\sum_{h \in H_g}\exp\left(\frac{I_{hg}}{1-\sigma_2}\right)
\end{equation}
The probability that consumer $i$ chooses group $g$ is given by:
\begin{equation}
P(g) = \Pr\Big\{ V_{ig} \geq V_{ig'} \quad \forall \; g'\neq g\Big\}
\end{equation}
Substituting the expression for $V_{ig}$ and simplifying:
\begin{equation}
P(g)=\Pr\left\{\zeta^g_{i} \geq I_{g'} - I_{g} +\zeta^{g'}_{i}\quad \forall g' \neq g\right\}
\end{equation}
Given that $\zeta^g_{i}$ follows a Type-I extreme value distribution, this choice probability resolves to the standard logit form:
\begin{equation}
P(g) = \frac{\exp(I_{g})}{\sum_{g'} \exp(I_{g'})}
\end{equation}
Substituting the expression for the group-level inclusive value:
\begin{equation}
P(g) =\frac{\left(\sum_{h \in H_g} \exp\left(\frac{I_{hg}}{1-\sigma_2}\right)\right)^{(1-\sigma_2)}}{\sum_{g'} \left(\sum_{h \in H_{g'}} \exp\left(\frac{I_{hg'}}{1-\sigma_2}\right)\right)^{(1-\sigma_2)}}
\end{equation}
Incorporating the complete hierarchical structure through the nested inclusive values yields the fully specified sequential logit probability:
\begin{equation}
\label{eq:group}
P(g) = \frac{\left(\sum_{h \in H_g} \left(\sum_{j \in J_{hg}} \exp\left(\frac{\delta_{j}}{1-\sigma_1}\right)\right)^{\frac{1-\sigma_1}{1-\sigma_2}}\right)^{(1-\sigma_2)}}{\sum_{g'}\left(\sum_{h \in H_{g'}} \left(\sum_{j \in J_{hg'}} \exp\left(\frac{\delta_{j}}{1-\sigma_1}\right)\right)^{\frac{1-\sigma_1}{1-\sigma_2}}\right)^{(1-\sigma_2)}}
\end{equation}
This equation characterizes the probability of selecting a product group under the sequential decision process, with nesting parameters $\sigma_1$ and $\sigma_2$ governing the information structure at each decision stage. The nested structure of inclusive values captures how consumers aggregate information across the entire choice hierarchy when making top-level decisions.

\subsection{Joint Choice Probability and \cite{berry1994estimating} Inversion}

The sequential decision structure yields a natural factorization of the joint probability into a product of conditional probabilities. For consumer $i$ selecting product $j$ in subgroup $h$ of group $g$, the joint probability is expressed as:
\begin{equation}
\label{eq:share}
s_{jhg} = P(j,h,g) = P(j|h,g) \cdot P(h|g) \cdot P(g)
\end{equation}
where $s_{jhg}$ represents the market share of product $j$ in subgroup $h$ of group $g$.\\
\newline
The derivation of product demand equations proceeds by inverting this market share equation to express mean utility as a function of observed market shares. By normalizing the inclusive value of the outside option (indexed as $g=0$) to zero ($I_{0} = 0$), the market share of the outside option becomes:
\begin{equation}
    s_{0}=P(g=0)=\frac{1}{1 + \sum_{g' \neq 0} \exp(I_{g'})}
\end{equation}
The ratio of product share to outside option share can be expressed by combining the previously derived choice probabilities:
\begin{equation}
\begin{aligned}
\frac{s_{jhg}}{s_{0}} =& \frac{\exp\left(\frac{\delta_{j}}{1-\sigma_1}\right)}{\sum_{j' \in J_{hg}} \exp\left(\frac{\delta_{j'}}{1-\sigma_1}\right)} \cdot \frac{\left(\sum_{j \in J_{hg}} \exp\left(\frac{\delta_{j}}{1-\sigma_1}\right)\right)^{\frac{1-\sigma_1}{1-\sigma2}}}{\sum_{h'\in H_g} \left(\sum_{j \in J_{h'g}} \exp\left(\frac{\delta_{j}}{1-\sigma_1}\right)\right)^{\frac{1-\sigma_1}{1-\sigma2}}} \cdot \\
&\left(\sum_{h\in H_g}  \left(\sum_{j \in J_{hg}} \exp\left(\frac{\delta_{j}}{1-\sigma_1}\right)\right)^{\frac{1-\sigma_1}{1-\sigma_2}}\right)^{(1-\sigma2)}
\end{aligned}
\end{equation}
Through rearrangement, simplification, and logarithmic transformation, this expression reduces to:
\begin{equation}
    \log\frac{s_{jhg}}{s_{0}}=\delta_{j}+\sigma_1\log\Pr(j|h,g)+\sigma_2\log\Pr(h|g)
\end{equation}
Using the conditional share notation, I arrive at the well known \cite{berry1994estimating} inversion formula:
\begin{equation}
\log\left(\frac{s_{jhg}}{s_{0}}\right) = \delta_{j}+\sigma_1\log s_{j|hg}+\sigma_2\log s_{h|g}
\end{equation}
where $s_{j|hg}$ represents the share of product $j$ within subgroup $h$ of group $g$, and $s_{h|g}$ represents the share of subgroup $h$ within group $g$.\\
\newline
This final expression is the two-level \cite{berry1994estimating} inversion formula, which provides a tractable linear relationship between observed market shares and the mean utility $\delta_j$. Remarkably, despite the distinct conceptual foundations of the sequential logit model, this inversion formula is identical to that derived for the two-level nested logit framework. 

\section{First Order Derivatives with Respect to $\delta_k$}
Taking the first order derivative with respect to $\delta_k$ of equation \eqref{eq:share} and keeping the share notation yields
\begin{equation}
    \frac{\partial s_{jhg}}{\partial\delta_k}=\frac{\partial s_{j|hg}}{\partial \delta_k}s_{h|g}s_g+s_{j|hg}\frac{\partial s_{h|g}}{\partial \delta_k}s_g + s_{j|hg}s_{h|g}\frac{\partial s_g}{\partial \delta_k}
\end{equation}

\subsection{Product Choice}
From equation \eqref{eq:product_choice} and using the share notation, let
\begin{equation}
    D = \sum_{j' \in J_{hg}} \exp\left(\frac{\delta_{j'}}{1-\sigma_1}\right)
\end{equation}
For $k=j$
\begin{equation}
\begin{aligned}
    \frac{\partial s_{j|hg}}{\partial \delta_j}
&= \frac{\frac{1}{1-\sigma_1}\exp\left(\frac{\delta_j}{1-\sigma_1}\right)\cdot D - \exp\left(\frac{\delta_j}{1-\sigma_1}\right)\cdot \frac{1}{1-\sigma_1}\exp\left(\frac{\delta_j}{1-\sigma_1}\right)}{D^2}\\
&=\frac{\frac{\exp\left(\frac{\delta_j}{1-\sigma_1}\right)}{1-\sigma_1}\left(D - \exp\left(\frac{\delta_j}{1-\sigma_1}\right)\right)}{D^2}\\
&=\frac{1}{1-\sigma_1}\frac{\exp\left(\frac{\delta_j}{1-\sigma_1}\right)}{D}\frac{\left(D - \exp\left(\frac{\delta_j}{1-\sigma_1}\right)\right)}{D}\\
&= \frac{1}{1-\sigma_1} s_{j|hg}\left[1 -  s_{j|hg}\right]
\end{aligned}
\end{equation}
For $k\neq j$, $k\in J_{hg}$
\begin{equation}
    \begin{aligned}
        \frac{\partial s_{j|hg}}{\partial \delta_k}
&= \frac{-\exp\left(\frac{\delta_j}{1-\sigma_1}\right)\cdot \frac{1}{1-\sigma_1}\exp\left(\frac{\delta_k}{1-\sigma_1}\right)}{D^2}\\
&= -\frac{1}{1-\sigma_1}\frac{\exp\left(\frac{\delta_j}{1-\sigma_1}\right)}{D}\frac{\exp\left(\frac{\delta_k}{1-\sigma_1}\right)}{D}\\
&=-\frac{1}{1-\sigma_1}s_{j|hg}s_{k|hg}
    \end{aligned}
\end{equation}
Expressed in a compact way
\begin{equation}
    \frac{\partial s_{j|hg}}{\partial \delta_k} =
\begin{cases}
\frac{1}{1-\sigma_1} s_{j|hg}\left[1 - s_{j|hg}\right], & \text{if } k = j \\[10pt]
-\frac{1}{1-\sigma_1} s_{j|hg}s_{k|hg}, & \text{if } k \neq j,\;k\in J_{hg} \\[10pt]
0, & \text{if } k\notin J_{hg}
\end{cases}
\end{equation}

\subsection{Subgroup Choice}
From equation \eqref{eq:subgroup} and using the share notation, let 
\begin{align}
\alpha &= \frac{1}{1-\sigma_1} \\
\beta &= \frac{1-\sigma_1}{1-\sigma_2} \\
S_h &= \sum_{j \in J_{hg}} \exp(\alpha\delta_j) \\
Z &= \sum_{h' \in H_g} S_{h'}^{\beta}
\end{align}
With this notation, I can rewrite 
\begin{align}
s_{h|g} = \frac{S_h^{\beta}}{Z}
\end{align}
For $k\in J_{hg}$
\begin{align}
\frac{\partial s_{h|g}}{\partial \delta_k} &= \frac{\frac{\partial}{\partial \delta_k}(S_h^{\beta}) \cdot Z - S_h^{\beta} \cdot \frac{\partial Z}{\partial \delta_k}}{Z^2}
\end{align}
Notice that
\begin{equation}
    \begin{aligned}
        \frac{\partial S_h}{\partial \delta_k} &= \frac{\partial}{\partial \delta_k}\left(\sum_{j \in J_{hg}} \exp(\alpha\delta_j)\right) \\
&= \frac{\partial}{\partial \delta_k}\left(\exp(\alpha\delta_k) + \sum_{j \in J_{hg}, j \neq k} \exp(\alpha\delta_j)\right) \\
&= \alpha \exp(\alpha\delta_k)
    \end{aligned}
\end{equation}
Then
\begin{equation}
    \begin{aligned}
        \frac{\partial}{\partial \delta_k}(S_h^{\beta}) &= \beta S_h^{\beta-1} \cdot \frac{\partial S_h}{\partial \delta_k} \\
&= \beta S_h^{\beta-1} \cdot \alpha \exp(\alpha\delta_k)
    \end{aligned}
\end{equation}
and
\begin{equation}
    \begin{aligned}
        \frac{\partial Z}{\partial \delta_k} &= \frac{\partial}{\partial \delta_k}\left(\sum_{h' \in H_g} S_{h'}^{\beta}\right) \\
&= \frac{\partial}{\partial \delta_k}(S_h^{\beta}) + \frac{\partial}{\partial \delta_k}\left(\sum_{h' \in H_g, h' \neq h} S_{h'}^{\beta}\right) \\
&= \beta S_h^{\beta-1} \cdot \alpha \exp(\alpha\delta_k)
    \end{aligned}
\end{equation}
Then, finally
\begin{equation}
    \begin{aligned}
        \frac{\partial s_{h|g}}{\partial \delta_k} &= \frac{\beta S_h^{\beta-1} \cdot \alpha \exp(\alpha\delta_k) \cdot Z - S_h^{\beta} \cdot \beta S_h^{\beta-1} \cdot \alpha \exp(\alpha\delta_k)}{Z^2} \\
&= \frac{\beta S_h^{\beta-1} \cdot \alpha \exp(\alpha\delta_k) \cdot (Z - S_h^{\beta})}{Z^2} \\
&= \beta \cdot \alpha \cdot \frac{S_h^{\beta-1} \cdot \exp(\alpha\delta_k)}{Z} \cdot \frac{Z - S_h^{\beta}}{Z} \\
&= \beta \cdot \alpha \cdot \frac{S_h^{\beta}}{Z} \cdot \frac{\exp(\alpha\delta_k)}{S_h} \cdot \left(1 - \frac{S_h^{\beta}}{Z}\right) \\
&= \beta \cdot \alpha \cdot s_{h|g} \cdot s_{k|h,g} \cdot (1 - s_{h|g}) \\
    \end{aligned}
\end{equation}
Which simplify in 
\begin{equation}
    \frac{\partial s_{h|g}}{\partial \delta_k}= \frac{1}{1-\sigma_2} \cdot s_{h|g} \cdot s_{k|h,g} \cdot (1 - s_{h|g})
\end{equation}
For $k\notin J_{hg}$ and $k\in J_{h'g}$ for some $h'\in H_g$
First,
\begin{align}
\frac{\partial S_h}{\partial \delta_k} &= 0
\end{align}
notice that for $h'$ where $k\in J_{h'g}$
\begin{align}
\frac{\partial S_{h'}}{\partial \delta_k} &= \alpha \exp(\alpha\delta_k)
\end{align}
Taking the derivative of $Z$
\begin{equation}
    \begin{aligned}
        \frac{\partial Z}{\partial \delta_k} &= \frac{\partial}{\partial \delta_k}\left(\sum_{h'' \in H_g} S_{h''}^{\beta}\right) \\
&= \frac{\partial}{\partial \delta_k}(S_{h'}^{\beta}) \\
&= \beta S_{h'}^{\beta-1} \cdot \frac{\partial S_{h'}}{\partial \delta_k} \\
&= \beta S_{h'}^{\beta-1} \cdot \alpha \exp(\alpha\delta_k)
    \end{aligned}
\end{equation}
Then
\begin{equation}
    \begin{aligned}
\frac{\partial s_{h|g}}{\partial \delta_k} &= \frac{\frac{\partial}{\partial \delta_k}(S_h^{\beta}) \cdot Z - S_h^{\beta} \cdot \frac{\partial Z}{\partial \delta_k}}{Z^2} \\
&= \frac{0 \cdot Z - S_h^{\beta} \cdot \beta S_{h'}^{\beta-1} \cdot \alpha \exp(\alpha\delta_k)}{Z^2} \\
&= -\frac{S_h^{\beta} \cdot \beta S_{h'}^{\beta-1} \cdot \alpha \exp(\alpha\delta_k)}{Z^2}
\end{aligned}
\end{equation}
Since
\begin{equation}
    \begin{aligned}
\frac{S_{h'}^{\beta-1} \cdot \exp(\alpha\delta_k)}{Z} &= \frac{S_{h'}^{\beta}}{Z} \cdot \frac{\exp(\alpha\delta_k)}{S_{h'}} \\
&= s_{h'|g} \cdot s_{k|h',g}
\end{aligned}
\end{equation}
Then, finally
\begin{equation}
    \begin{aligned}
\frac{\partial s_{h|g}}{\partial \delta_k} &= -\beta \cdot \alpha \cdot s_{h|g} \cdot s_{h'|g} \cdot s_{k|h',g} \\
&= -\frac{1}{1-\sigma_2} \cdot s_{h|g} \cdot s_{h'|g} \cdot s_{k|h',g}
\end{aligned}
\end{equation}
For $k\notin J_{hg}$ and $k\in J_{h'g}$ for some $h'\notin H_g$
\begin{align}
\frac{\partial s_{h|g}}{\partial \delta_k} &= 0
\end{align}
Expressed in a compact way
\begin{equation}
    \frac{\partial s_{h|g}}{\partial \delta_k} =
\begin{cases}
\frac{1}{1-\sigma_2} \cdot s_{h|g} \cdot s_{k|h,g} \cdot (1 - s_{h|g}), & \text{if } k \in J_{hg} \\[10pt]
-\frac{1}{1-\sigma_2} \cdot s_{h|g} \cdot s_{h'|g} \cdot s_{k|h',g}, & \text{if } k\notin J_{hg} \text{ and } k\in J_{h'g} \text{ for some } h'\in H_g \\[10pt]
0, & \text{if } k\notin J_{hg} \text{ and } k\in J_{h'g} \text{ for some } h'\notin H_g
\end{cases}
\end{equation}

\subsection{Group Choice}
Let
\begin{align}
\alpha &= \frac{1}{1-\sigma_1} \\
\beta &= \frac{1-\sigma_1}{1-\sigma_2} \\
\gamma &= 1-\sigma_2 \\
S_{hg} &= \sum_{j \in J_{hg}} \exp(\alpha\delta_j) \\
T_g &= \sum_{h \in H_g} S_{hg}^{\beta} \\
Z &= \sum_{g'} T_{g'}^{\gamma}
\end{align}
With this notation, one can rewrite equation \eqref{eq:group} using shares as
\begin{align}
s_g = \frac{T_g^{\gamma}}{Z}
\end{align}
For $k\in J_{hg}$ and $h\in H_g$, first, calculating the derivative of $S_{hg}$ with respect to $\delta_k$:
\begin{equation}
   \begin{aligned}
\frac{\partial S_{hg}}{\partial \delta_k} &= \frac{\partial}{\partial \delta_k}\left(\sum_{j \in J_{hg}} \exp(\alpha\delta_j)\right) \\
&= \alpha \exp(\alpha \delta_k)
\end{aligned} 
\end{equation}
Next, the derivative of $T_g$ with respect to $\delta_k$:
\begin{equation}
    \begin{aligned}
        \frac{\partial T_g}{\partial \delta_k} &= \frac{\partial}{\partial \delta_k}\left(\sum_{h' \in H_g} S_{h'g}^{\beta}\right) \\
&= \frac{\partial}{\partial \delta_k}(S_{hg}^{\beta}) \\
&= \beta S_{hg}^{\beta-1} \cdot \frac{\partial S_{hg}}{\partial \delta_k} \\
&= \beta S_{hg}^{\beta-1} \cdot \alpha \exp(\alpha \delta_k)
    \end{aligned}
\end{equation}
And the derivative of $T_g^\gamma$ with respect to $\delta_k$:
\begin{equation}
    \begin{aligned}
        \frac{\partial T_g^{\gamma}}{\partial \delta_k} &= \gamma T_g^{\gamma-1} \cdot \frac{\partial T_g}{\partial \delta_k} \\
&= \gamma T_g^{\gamma-1} \cdot \beta S_{hg}^{\beta-1} \cdot \alpha \exp(\alpha \delta_k)
    \end{aligned}
\end{equation}
The derivative of $Z$ with respect to $\delta_k$:
\begin{equation}
    \begin{aligned}
        \frac{\partial Z}{\partial \delta_k} &= \frac{\partial}{\partial \delta_k}\left(\sum_{g'} T_{g'}^{\gamma}\right) \\
&= \frac{\partial T_g^{\gamma}}{\partial \delta_k} \\
&= \gamma T_g^{\gamma-1} \cdot \beta S_{hg}^{\beta-1} \cdot \alpha \exp(\alpha \delta_k)
    \end{aligned}
\end{equation}
Then
\begin{equation}
    \begin{aligned}
        \frac{\partial s_g}{\partial \delta_k} &= \frac{\frac{\partial T_g^{\gamma}}{\partial \delta_k} \cdot Z - T_g^{\gamma} \cdot \frac{\partial Z}{\partial \delta_k}}{Z^2} \\
&= \frac{\gamma T_g^{\gamma-1} \cdot \beta S_{hg}^{\beta-1} \cdot \alpha \exp(\alpha \delta_k) \cdot Z - T_g^{\gamma} \cdot \gamma T_g^{\gamma-1} \cdot \beta S_{hg}^{\beta-1} \cdot \alpha \exp(\alpha \delta_k)}{Z^2} \\
&= \frac{\gamma \beta \alpha T_g^{\gamma-1} S_{hg}^{\beta-1} \exp(\alpha \delta_k) (Z - T_g^{\gamma})}{Z^2} \\
&= \gamma \beta \alpha \cdot \frac{T_g^{\gamma-1} S_{hg}^{\beta-1} \exp(\alpha \delta_k)}{Z} \cdot \frac{Z - T_g^{\gamma}}{Z}\\
&= \gamma \beta \alpha \cdot \frac{T_g^{\gamma}}{Z} \cdot \frac{1}{T_g} \cdot \frac{S_{hg}^{\beta}}{T_g} \cdot \frac{1}{S_{hg}} \cdot \frac{\exp(\alpha \delta_k)}{S_{hg}} \cdot S_{hg} \cdot \left(1 - \frac{T_g^{\gamma}}{Z}\right) \\
&= \gamma \beta \alpha \cdot s_g \cdot s_{h|g} \cdot s_{k|hg} \cdot (1 - s_g)\\
&= s_g \cdot s_{h|g} \cdot s_{k|hg} \cdot (1 - s_g)\\
& = s_{khg}(1-s_g)
    \end{aligned}
\end{equation}
By the law of total probability and since
\begin{equation}
    \begin{aligned}
        s_g &= \frac{T_g^{\gamma}}{Z} \\
s_{h|g} &= \frac{S_{hg}^{\beta}}{T_g} \\
s_{k|hg} &= \frac{\exp(\alpha \delta_k)}{S_{hg}}
    \end{aligned}
\end{equation}
For $k\in J_{h'g'}$ and $h'\in H_{g'}$ with $g'\neq g$, for $\delta_k$ affecting a diferent group:
\begin{align}
\frac{\partial S_{h'g'}}{\partial \delta_k} &= \alpha \exp(\alpha \delta_k) \\
\frac{\partial T_{g'}}{\partial \delta_k} &= \beta S_{h'g'}^{\beta-1} \cdot \alpha \exp(\alpha \delta_k) \\
\frac{\partial T_{g'}^{\gamma}}{\partial \delta_k} &= \gamma T_{g'}^{\gamma-1} \cdot \beta S_{h'g'}^{\beta-1} \cdot \alpha \exp(\alpha \delta_k)
\end{align}
Since $\delta_k$ does not affect the numerator of $s_g$:
\begin{equation}
    \begin{aligned}
        \frac{\partial s_g}{\partial \delta_k} &= \frac{0 \cdot Z - T_g^{\gamma} \cdot \frac{\partial Z}{\partial \delta_k}}{Z^2} \\
&= -\frac{T_g^{\gamma} \cdot \gamma T_{g'}^{\gamma-1} \cdot \beta S_{h'g'}^{\beta-1} \cdot \alpha \exp(\alpha \delta_k)}{Z^2} \\
&= -\gamma \beta \alpha \cdot \frac{T_g^{\gamma}}{Z} \cdot \frac{T_{g'}^{\gamma-1} \cdot S_{h'g'}^{\beta-1} \cdot \exp(\alpha \delta_k)}{Z}
    \end{aligned}
\end{equation}
Expressing in terms of probabilities
\begin{equation}
    \begin{aligned}
        \frac{T_{g'}^{\gamma-1} \cdot S_{h'g'}^{\beta-1} \cdot \exp(\alpha \delta_k)}{Z} &= \frac{T_{g'}^{\gamma}}{Z} \cdot \frac{1}{T_{g'}} \cdot \frac{S_{h'g'}^{\beta}}{T_{g'}} \cdot \frac{1}{S_{h'g'}} \cdot \frac{\exp(\alpha \delta_k)}{S_{h'g'}} \cdot S_{h'g'} \\
&= s_{g'} \cdot s_{h'|g'} \cdot s_{k|h'g'}
    \end{aligned}
\end{equation}
And then 
\begin{equation}
    \begin{aligned}
        \frac{\partial s_g}{\partial \delta_k} &= -\gamma \beta \alpha \cdot s_g \cdot s_{g'} \cdot s_{h'|g'} \cdot s_{k|h'g'} \\
&= -(1-\sigma_2) \cdot \frac{1-\sigma_1}{1-\sigma_2} \cdot \frac{1}{1-\sigma_1} \cdot s_g \cdot s_{g'} \cdot s_{h'|g'} \cdot s_{k|h'g'} \\
&= -s_g \cdot s_{g'} \cdot s_{h'|g'} \cdot s_{k|h'g'} \\
&= -s_g \cdot s_{kh'g'} \\
    \end{aligned}
\end{equation}
Expressed in a compact way
\begin{equation}
    \frac{\partial s_{g}}{\partial \delta_k} =
\begin{cases}
s_{khg}(1-s_g), & \text{if } k \in J_{h'g} \text{ and } h'\in H_g\\[10pt]
-s_g \cdot s_{kh'g'} , & \text{if } k\in J_{h'g'} \text{ and } h'\in H_{g'} \text{ for some }g'\neq g 
\end{cases}
\end{equation}

\clearpage
\bibliography{\bib}

\end{document}